\documentclass[aps,pre,twocolumn,groupedaddress,showpacs,amsfonts,10pt,%
tightenlines,floatfix]{revtex4-1}

\usepackage[utf8]{inputenc}
\usepackage{graphicx}
\usepackage{amsmath,amssymb}
\usepackage{upgreek}
\usepackage[usenames,dvipsnames]{xcolor}
\usepackage{braket}
\usepackage{cancel}

\graphicspath{{./Abbildungen/}}

\begin{document}

\title{Critical excitation-inhibition balance in dense neural networks}

\author{Lorenz Baumgarten}
\email[]{lbaumgarten@itp.uni-bremen.de}

\author{Stefan Bornholdt}
\email[]{bornholdt@itp.uni-bremen.de}

\affiliation{Institut f\"ur Theoretische Physik, Universit\"at Bremen, 28759 Bremen, Germany}

\date{\today}

\begin{abstract}
The ``edge of chaos'' phase transition in artificial neural networks is of
renewed interest in light of recent evidence for criticality in brain dynamics.
Statistical mechanics traditionally studied this transition with connectivity $k$
as the control parameter and an exactly balanced excitation-inhibition ratio.
While critical connectivity has been found to be low in these model systems,
typically around $k=2$, which is unrealistic for natural neural systems,
a recent study utilizing the excitation-inhibition ratio as the control parameter
found a new, nearly degree independent, critical point when connectivity is large.
However, the new phase transition is accompanied by an unnaturally high level
of activity in the network.

Here we study random neural networks with the additional properties of
(i) a high clustering coefficient and (ii) neurons that are solely either excitatory
or inhibitory, a prominent property of natural neurons.
As a result we observe an additional critical point for networks with large connectivity,
regardless of degree distribution,
which exhibits low activity levels that compare well with neuronal brain networks.
\end{abstract}

\maketitle

\begin{figure}
\includegraphics[width=1\linewidth]{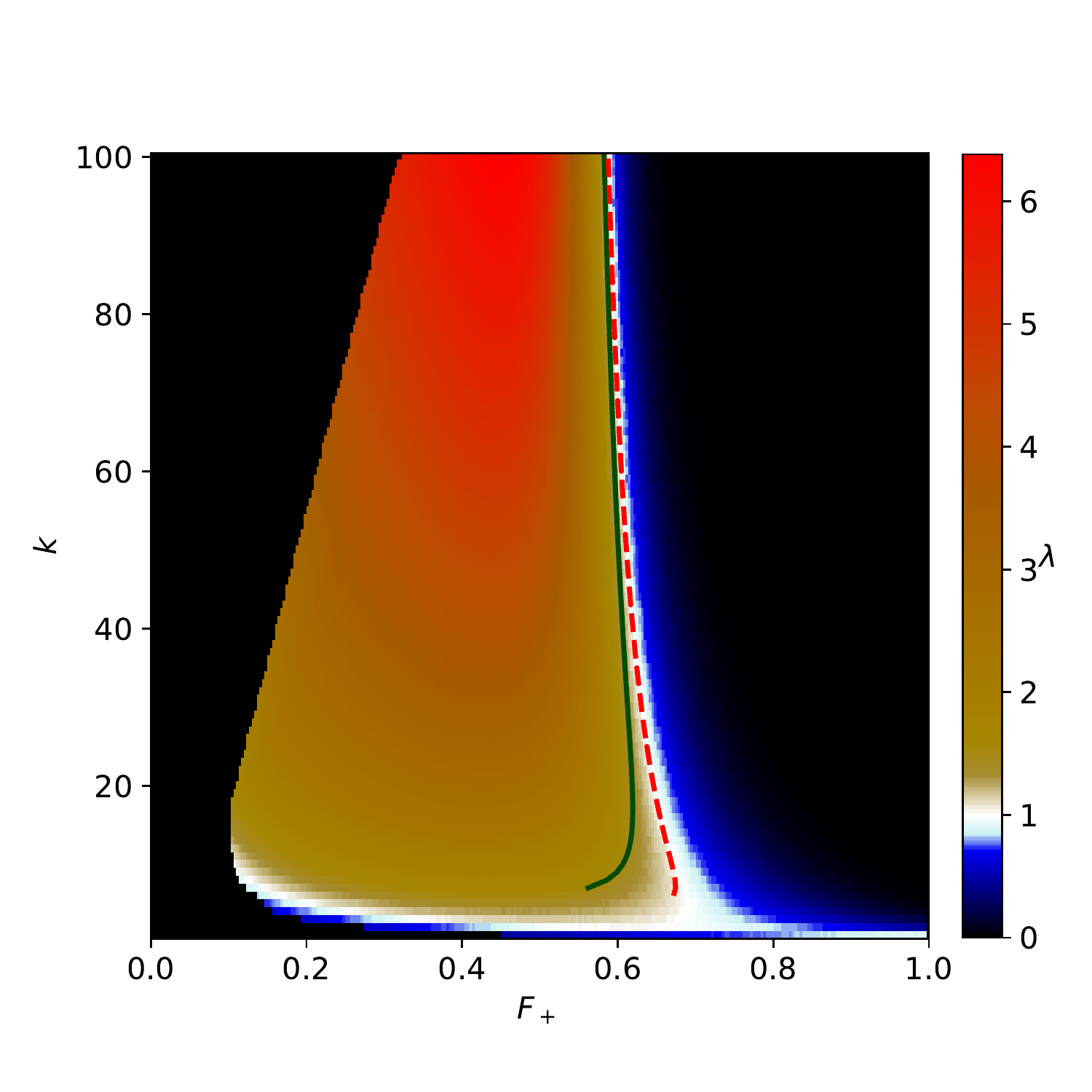}
\caption{Sensitivity as a function of fraction of excitatory links $F_+$ and connectivity $k$
in random neural networks, similar to Fig.\ 1 C in \cite{BornholdtNeto2017}, 
for threshold $h=0$ and $N=10^3$ nodes. 
Lines compare Eq.\ \eqref{F_+(k)} (green solid line)
and the numerical solution of Eq.\ \eqref{p_s^+(k)} (red dashed line)
with the simulation results.
Both lines approximate the simulation's critical line well for large $k$
Note that the left flank of the sensitive region of the simulation does not 
exhibit a (white) critical corridor which is further discussed in the text.}
\label{ParamScanApprox}
\end{figure}

Between the ordered and chaotic regimes of threshold neural networks lies
the ``edge of chaos,'' a critical point where the length and size distributions
of activity avalanches are governed by characteristic power laws.
This dynamical phase transition has been thoroughly studied in random neural networks
\cite{Kuerten1988,BornholdtRohlf2002,Szejka2008,Rohlf2008},
non symmetric spin glasses \cite{Derrida1987}, and random Boolean networks
\cite{Derrida1986,Kuerten1988b,AldanaCoppersmithKadanoff2003,RandomBooleanNetworksDrossel2008,BornholdtKauffman2019}.
Traditionally, threshold neural networks have been studied 
with precisely balanced excitation and inhibition, usually by randomly
assigning activating and inhibiting links with equal probabilities. 
In these networks, criticality occurs for small average degrees $k$ \cite{Kuerten1988}.
However, when allowing the fraction of excitatory links $F_+$ as a
second control parameter of the phase transition, it was recently discovered that
there exist two critical lines in the $k$-$F_+$-plane:
one almost parallel to the $F_+$ axis at low $k$ and one almost independent
of $k$ at some $F_+>0.5$ \cite{BornholdtNeto2017}, see Fig.\ \ref{ParamScanApprox}.
\\
The relevance of this new critical point becomes apparent in the context of neural brain networks
which exhibit a high average degree ($k\approx10^4$ in human brains \cite{Huttenlocher1979})
and a characteristic imbalance between excitation and inhibition
(\mbox{20--30\,\%} of neurons are inhibitory in monkey brains \cite{Hendry1987}).
There is a large amount of evidence suggesting that the brain operates near a critical point,
namely avalanche sizes and durations governed by power laws
\cite{Beggs2003, Friedman2012, Priesemann2014, Timme2016, Yaghoubi2018, Fontenele2019},
the possibility of tuning from a subcritical regime through the critical point to
a supercritical regime \cite{Beggs2005}, mathematical relations between critical exponents,
and collapsable avalanche shapes \cite{Friedman2012,Shaukat2016,Fontenele2019}.
Further, Fraiman {\em et al.}\ showed striking similarities between correlation networks
extracted from brains and the Ising model at the critical point \cite{Fraiman2009}.
The interest in the role of criticality in the brain is illustrated by the large amount of research
devoted to criticality in network models inspired by biological networks
\cite{Gross2008, Li2017, Clawson2017, Brochini2016, Gautam2015, Shriki2016, Rodriguez2017, Ferraz2017}.

Unfortunately, the high-degree critical point of Fig.\ \ref{ParamScanApprox} exists in 
a high-activity regime which is unrealistic for brain networks.
We find, however, an additional critical point that persists at low activities, at the left flank of 
the high sensitivity region, when including
additional network properties characteristic of brain networks, thereby providing
a more likely network model candidate for describing the processes behind brain criticality.

\begin{figure}
	\includegraphics[width=1\linewidth]{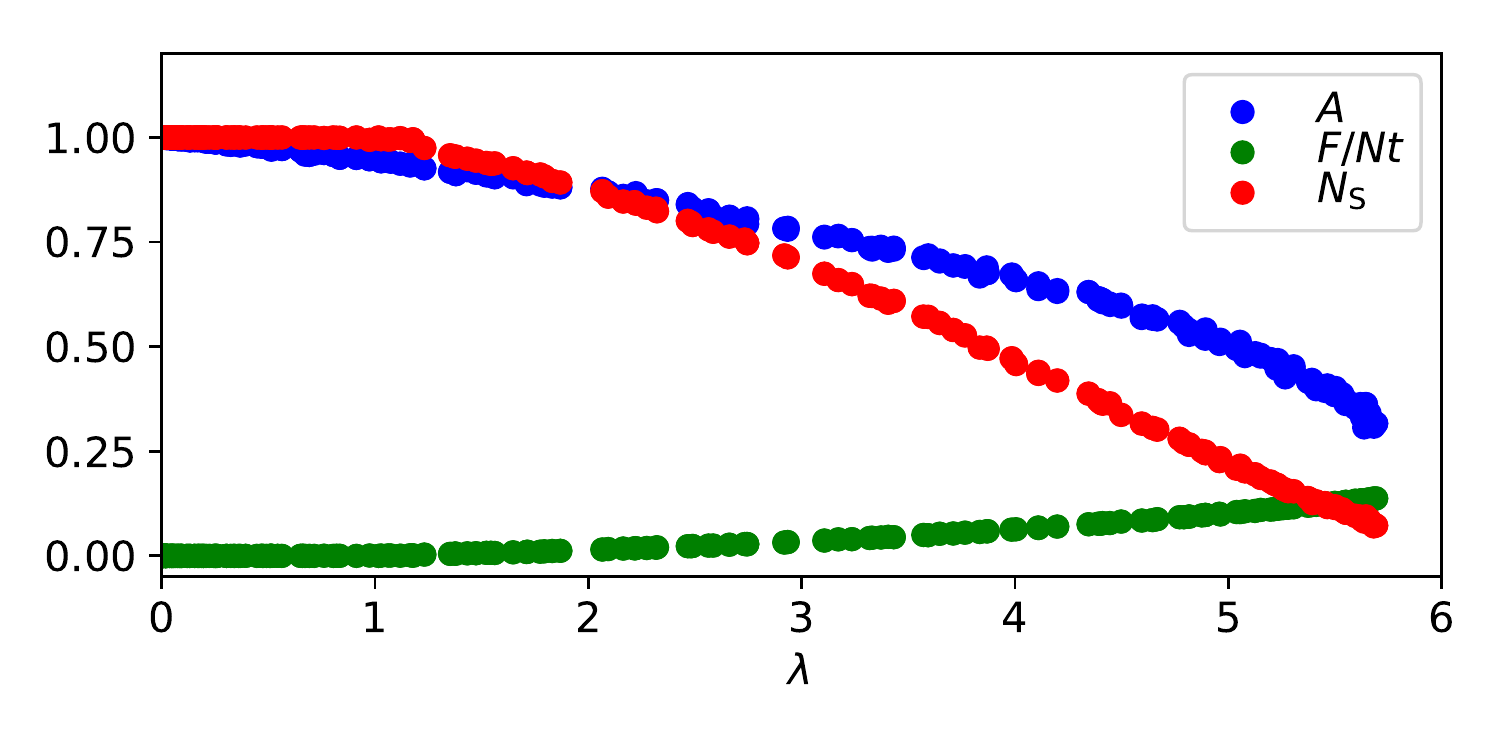}
	\caption{ Activity $A$, static node fraction $N_\text S$, and flips per node and time step $F/Nt$ as a function of the sensitivity $\lambda$ for $k=80$, $N=10^4$, and $h=1$. }
	\label{F-Measure_1_1e4_80}
\end{figure}

We use threshold networks consisting of $N$ nodes connected by $kN$ directed edges,
whose node states are updated in parallel according to
\begin{align}
\sigma_i(t+1) = \begin{cases}
		1 & \text{, if} \sum_{j=1}^N w_{ij}\sigma_j(t) > h\\
		0 & \text{, if} \sum_{j=1}^N w_{ij}\sigma_j(t) \le h,
		\end{cases}
\end{align}
where $\sigma_i(t)$ is the node $i$'s state at time $t$ and $w_{ij}$ is the weight of the connection from node $j$ to node $i$.
The weights $w_{ij}$ can be $0$ if there is no connection between nodes $i$ and $j$, or $\pm 1$ otherwise.
The weights of existing connections are chosen randomly with excitatory links $w_{ij}=+1$ chosen with probability $F_+$.
Initial states of the nodes are chosen according to a random initial activity $A_0 = \frac 1 N \sum_i \sigma_i$.
\\
A simple quantity that we use to measure criticality is the sensitivity $\lambda$ \cite{LuqueSole1997, ShmulevichKauffman2004}.
Imagine switching one node's state in the current time step; then $\lambda$ is defined as the average number of nodes
whose states will then be different in the next time step from what they would have been otherwise.
If sensitivity is smaller or larger than 1, perturbations will quickly die out or dominate the entire network, respectively.
Hence, at $\lambda=1$, the network is in a critical state.
\\
First, in order to establish whether the vertical white line defined by $\lambda=1$ seen in Fig.\  \ref{ParamScanApprox}
indeed is a critical point, we measure the averages of multiple quantities of interest, as well as
the average sensitivity for $10^3$ time steps after letting the network relax from its initial condition
within $2\cdot10^3$ time steps (tests show that increasing this time or waiting until an attractor is reached\,---\,where possible,
attractors cannot be found in a reasonable amount of time for $\lambda\gg1$\,---\,does not change the results)
for different values of $F_+$. Afterwards, we can plot the measured quantities as a function of  sensitivity.
The measured quantities are the network's activity $A$, the fraction of nodes which do not change their state
within the $10^3$ time steps $N_\text S$, and the average number of state changes per node and time step $F/Nt$.
This measurement is shown in Fig.\ \ref{F-Measure_1_1e4_80}.
\\
For $\lambda<1$, essentially all nodes are static 
(i.e.\ remaining in one state, either active or inactive) 
and almost no flips happen, whereas for $\lambda>1$ the number
of static nodes drops and the number of flips increases, so $\lambda=1$ is a boundary between order and chaos.
Also note that the network's activity is very high at the critical point. It seems, therefore, that this critical
point cannot underlie a mechanism that defines criticality in the brain, as almost all neurons constantly firing is not realistic.
\\
Further, we measure avalanche sizes and durations at the critical point, as described in the Supplemental Material
\cite{SuppMat}; see Fig.\ \ref{Histograms}.
\begin{figure}
	\includegraphics[width=1\linewidth]{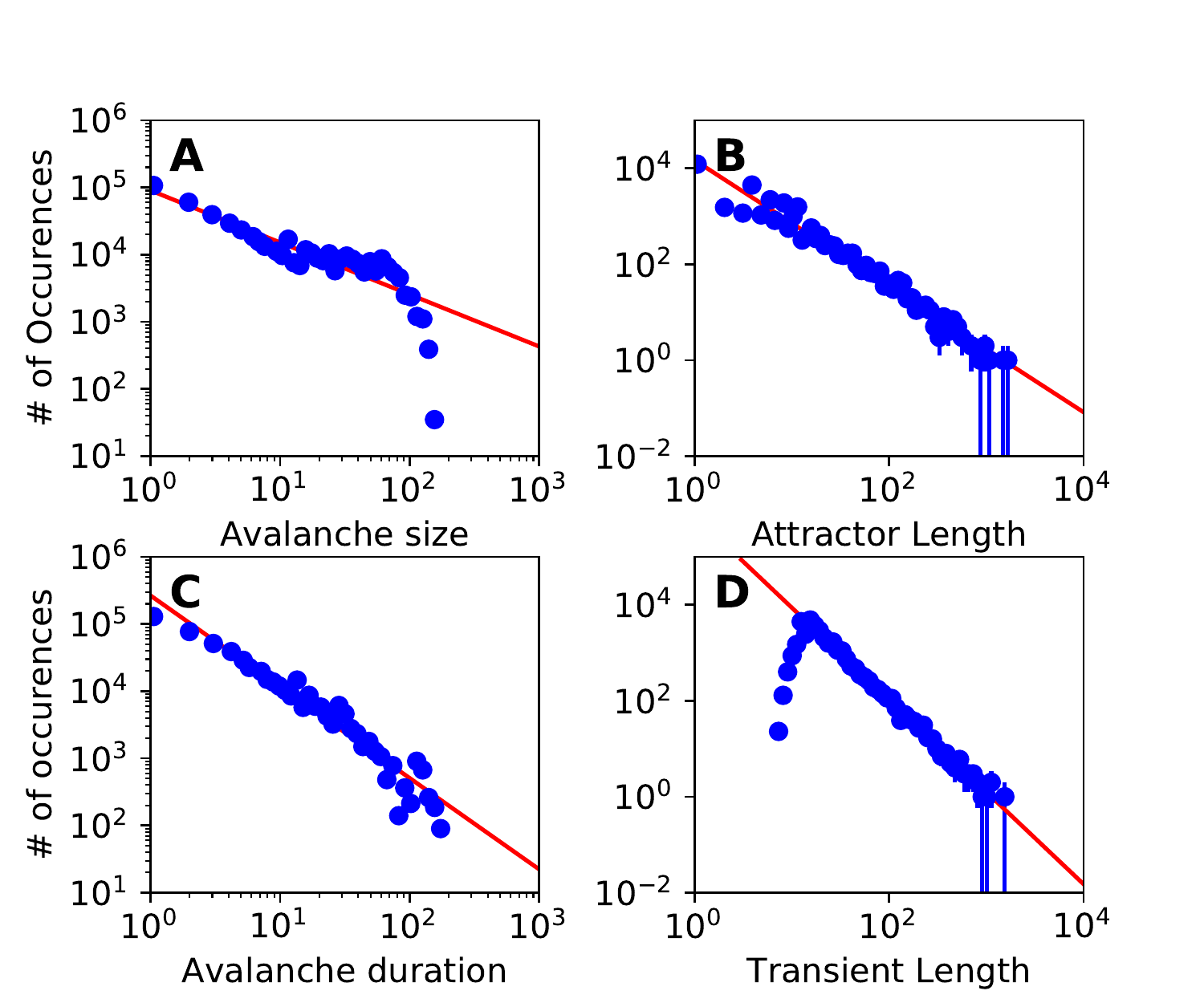}
	\caption{Distributions of avalanches.
	(a) Sizes and
	(c) durations for networks with $F_+\approx0.6$, $k=80$, $N=10^4$, and $h=1$. 
	The slopes shown in red are (a) $-0.8$ and (c) $-1.4$.
	\\
	Also,
	(b) attractor and
	(d) transient length distributions for networks with $0.95\le\lambda\le1.05$, $F_+\approx0.6$, $k=80$, $N=4444$, and $h=1$.
	The slopes shown in red are (b) $-1.3$ and (d) $-1.9$. Logarithmic binning is used for all four figures.}
	\label{Histograms}
\end{figure}
We observe power laws in both avalanche size and duration distributions.
\\
Finally, we measure the attractor and transient lengths, as well as the average sensitivity
within the attractor for a number of different network realizations for fixed parameters.
We only use parameter and attractor lengths of networks whose average sensitivity $\lambda$
is within $1-\delta\le\lambda\le1+\delta$ with $\delta=0.05$.
Attractor and transient length distributions are shown in FIG.\ \ref{Histograms}.
Both the attractor lengths as well as the right flank of the transient length distributions
show clear power laws, as is to be expected for critical networks \cite{Liang1996}.
\\
All of the above discussed properties lead us to conclude that this is indeed a critical point.
\\
We use Derrida's annealed approximation \cite{Derrida1986},
adopted for threshold networks \cite{BornholdtRohlf2002},
to estimate the critical $F_+$ as a function of $k$, and arrive at the equation
\begin{align}
\frac 1 k = \binom k {\frac{k+h+1}2} F_+^{\frac {k+h+1} 2}(1-F_+)^{\frac{k-h-1}2} \frac {k+h+1}{2k}.
\label{p_s^+(k)}
\end{align}
Under the assumption of large average degree $k\gg h$, $k\gg 1$, this can be simplified to
\begin{align}
F_+ &= \frac 1 2 \left[ 1 + \left\{1-\left(\frac{2\pi}k\right)^{\frac k 2}\right\}^\frac 1 2\right].
\label{F_+(k)}
\end{align}
See Supplemental Material \cite{SuppMat} for details.
Figure \ref{ParamScanApprox} shows a comparison of Eq.\ \eqref{F_+(k)},
as well as the numerical solution of Eq.\ \eqref{p_s^+(k)}, with our simulation results.

Let us now focus on the the left flank of the central high sensitivity 
region in Fig.\ \ref{ParamScanApprox}. When lowering the value of $F_+$ 
from intermediate values towards $0$, sensitivity $\lambda$ seems to 
suddenly drop to $0$ from values larger than $1$. In the simulations 
this is due to a sudden drop in persistent activity: All activity dies 
out before the average sensitivity crosses through one. 
Critical sensitivity here falls into the left (black) region of 
entirely inactive networks whose sensitivity is not shown (as only 
persisting activity is relevant and, therefore, plotted). 

However, as a central observation of our study, we find that networks 
can be kept from abruptly dying out for low $F_+$ by introducing 
two properties to the network: increasing the networks' clustering 
coefficient $C$ and requiring that nodes have either only excitatory 
or only inhibitory outgoing edges (Dale's principle). Both of these 
properties are prominent features of brain networks 
\cite{Hilgetag2000, Sporns2004, Sporns2006, Bullmore2006, Dale1935}.
Note that these properties do not necessarily cause networks to show 
finite activity for values of $F_+$ in which the random network has 
zero activity but instead that the activity goes continuously 
towards zero with lowering $F_+$ instead of abruptly dropping to zero. \\

We believe the mechanism underlying the left flank's survival to be as follows:
If two excitatory nodes which are connected to each other are active, 
then for high clustering coefficients, they are likely to have shared neighbors 
and can therefore combine their efforts to also activate these neighbors 
more easily than in random networks and thereby create islands of surviving activity.
The contribution of nodes being either excitatory or inhibitory 
is likely that if few random nodes are active within a region, 
this property significantly increases the variance of the  
relative number of activating signals in that region 
and therefore increases the probability of areas exhibiting 
high excitation by random chance.\\

\begin{figure}
	\includegraphics[width=1\linewidth]{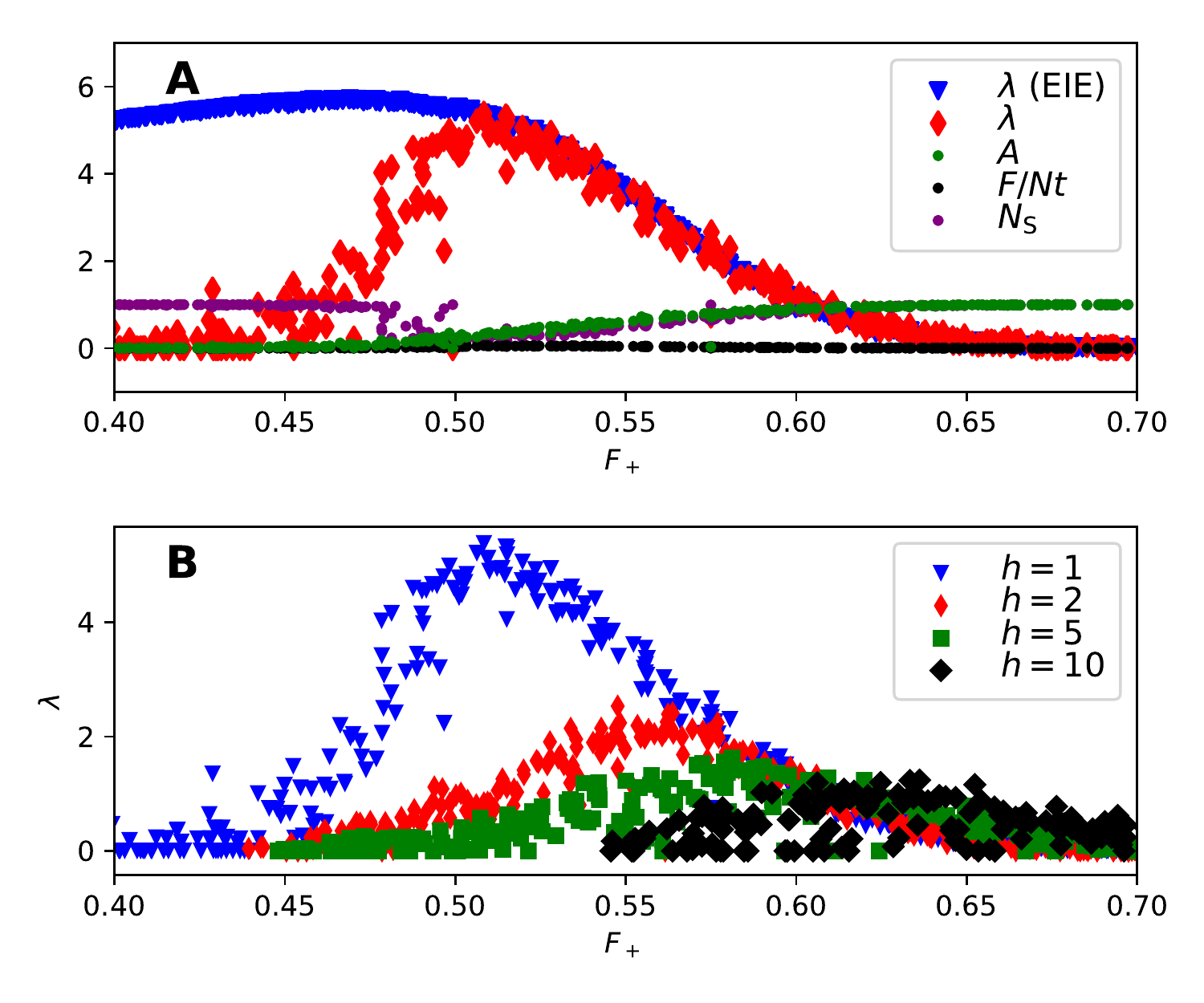}
	\caption{
	(a) Activity $A$, static node fraction $N_\text S$, flips per node and time step $F/Nt$, and sensitivity $\lambda$ at $h=1$ and
	(b) sensitivity $\lambda$ for different thresholds $h$ as a function of $F_+$ for $k=80$, $N=10^4$,
	and WS-EIN networks with rewiring probability $\beta=10^{-2}$ ($C\approx0.72$).
	The sensitivity for an equivalent ER network ($C=0.008$) is also shown in A for comparison.}
	\label{SecondCritPoint}
\end{figure}

We also see that the sensitivity in clustered graphs with nodes either 
fully excitatory or inhibitory closely follows the sensitivity of random graphs 
for high values of $F_+$ but then drops off for lower $F_+$, see Fig.\  
\ref{SecondCritPoint}.
This is likely due to nodes in the center of activity islands receiving 
many more excitatory connections than necessary for activation.
This both lowers the overall activity because these redundant excitatory 
signals essentially lower the network's total excitation and lower
the sensitivity because only nodes with an input sum near the  
excitation threshold contribute to it.\\

Networks with only a high clustering coefficient, without the second property
of nodes having either only excitatory or only inhibitory outgoing edges, 
can also show surviving activity on the left flank for some initial 
configurations and for exceedingly high clustering coefficients and thresholds, 
but even then the left flank drops sharply towards zero.
In the following, let us denote networks obeying Dale's principle \cite{Dale1935},
i.e.\ networks consisting of excitatory neurons and inhibitory neurons as 
{\em EIN networks}, as opposed to networks with excitatory/inhibitory edges 
which we will call {\em EIE networks}.

\begin{figure}
\includegraphics[width=1\linewidth]{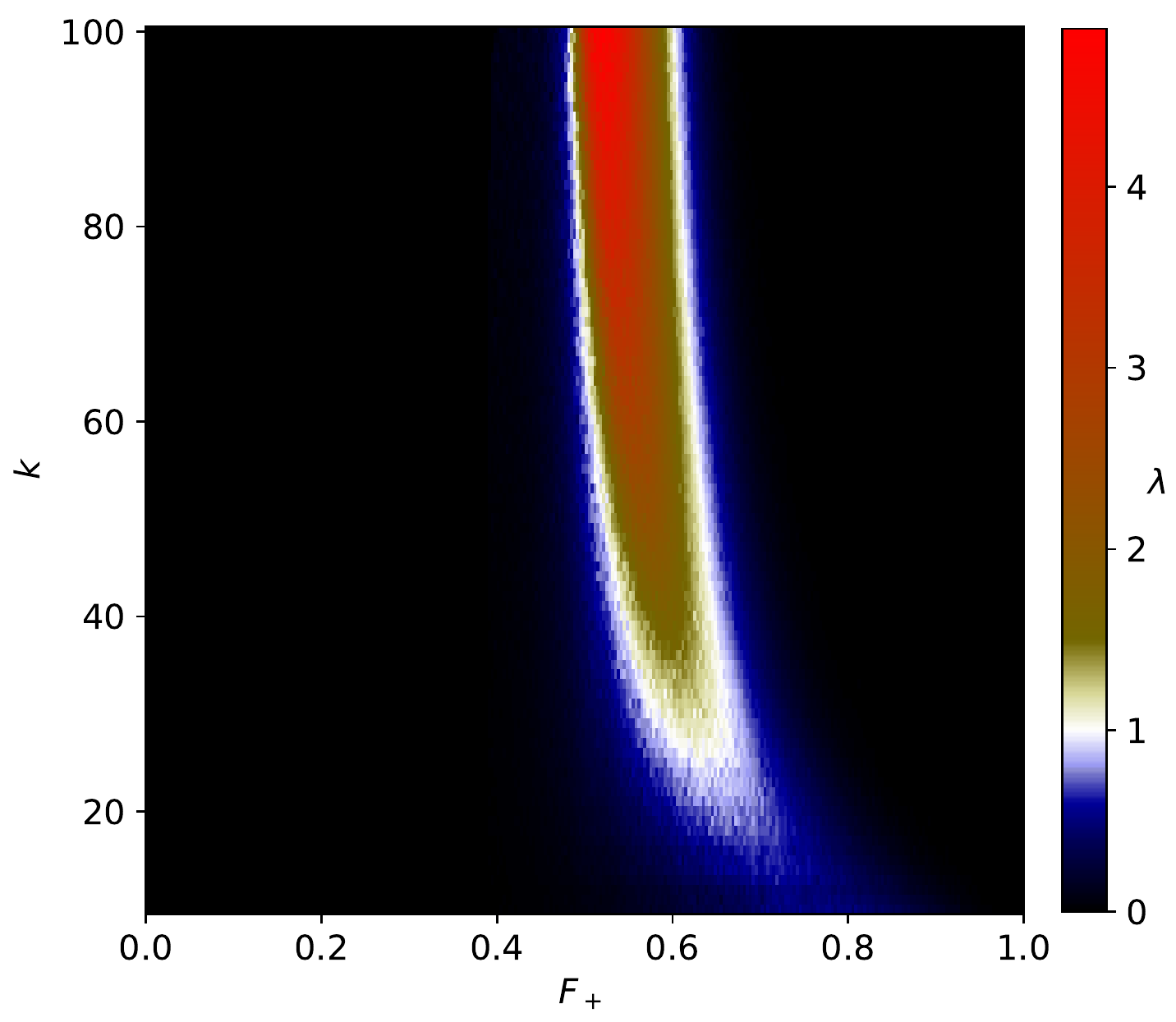}
\caption{Sensitivity as a function of fraction of excitatory links $F_+$ and connectivity $k$
in clustered EIN (Dale) neural networks for threshold $h=2$, clustering coefficient $C=0.65$, 
and $N=10^4$ nodes. 
}
\label{SingleParameterScan}
\end{figure}

Since the network's activity does not abruptly die out on the left flank 
anymore for clustered EIN networks, a second critical point can be found here, 
as shown in Fig.\ \ref{SingleParameterScan} and Fig.\ \ref{SecondCritPoint} (a). 
Plotting the sensitivity in the $F_+$-$k$-plane in Fig. \ref{SingleParameterScan}, 
we now see that the left flank indeed exhibits critical sensitivity $\lambda =1$ (white color). 
Note that, in contrast to the first critical point at the right flank of the sensitive 
region, this second critical point at the left flank exists in a low-activity state, 
making it more interesting for real life applications, such as studying mechanisms 
underlying brain criticality.

To construct networks with different high clustering coefficients, we here use 
directed Watts-Strogatz (WS) networks \cite{WattsStrogatz1998, Song2014}.
The original WS model consists of a ring of $N$ neurons with periodic boundary 
conditions in which every neuron is connected to its $k$ nearest neighbors. 
Then, connections are randomly rewired with rewiring probability $\beta$.
We use an essentially equivalent implementation without explicit rewiring 
from \cite{Song2014} in which the probability of a connection from a node $i$ 
to a node $j$ existing is
\begin{align}
	p_{ij} = &\beta p_0 + (1-\beta)\Theta[p_0-D_{ij}/(N/2)] \nonumber\\
	&+ \frac 1 2 (1-\beta)  \Theta[p_0 + D_{ij}/(N/2)] \nonumber\\
	& \times \Theta[p_0-(D_{ij}-0.5)/(N/2)],
\end{align}
where $p_0=k/(N-1)$ and $D_{ij}$ is the distance between nodes $i$ and $j$ on the ring, 
i.e. $D_{ij}=min(|i-j|,N-|i-j|)$.
The third term has been added to enable uneven values of $k$.
By manipulating the rewiring probability $\beta$, we can vary a network's clustering 
coefficient and average path length.
The Watts-Strogatz model's strength is that when varying $\beta$, there is a region 
in which the clustering coefficient is nearly constant while the average path length 
changes drastically and a second region in which the clustering coefficient
changes and the average path length is nearly constant, enabling us to isolate these 
two parameters' effects.

In our study of clustered EIN networks, we find that the second critical point comes 
into existence in the region in which the clustering coefficient changes,
while it is unaffected by changes within the region in which the clustering coefficient is constant.
Therefore, a high clustering coefficient is sufficient to enable the second critical point's existence.

\begin{figure*}
	\includegraphics[width=1\linewidth]{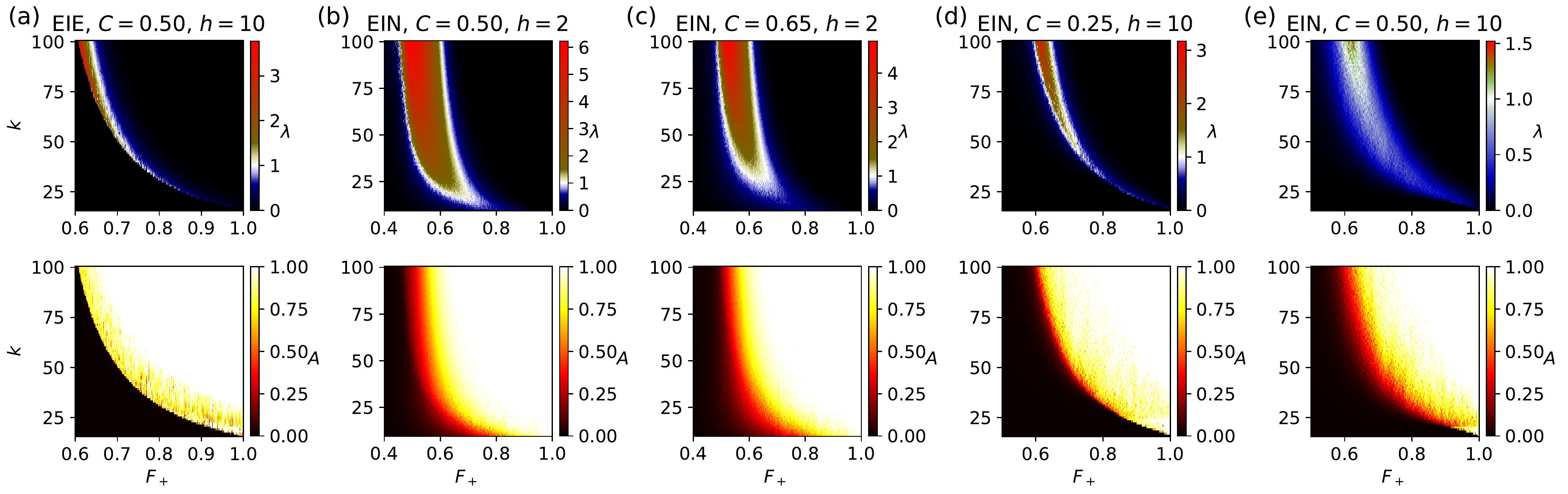}
	\caption{Sensitivity $\lambda$ and activity $A$ for different network configurations.
	White denotes a critical sensitivity.
	(a) For EIE networks\,---\,except for very high clustering coefficients and thresholds\,---\,the left flank
	dies out before reaching the critical point.
	(b) Switching to an EIN network stabilizes the left flank; however, it still collapses for high average degrees $k$
	without a high clustering coefficient. Some white artifacts can be seen because the left flank
	does not die out within $2\cdot10^3$ time steps; it does, however, die out after a larger number of time steps,
	and therefore no second critical point exists here, see Supplemental Material \cite{SuppMat} for more information.
	(c) A higher clustering coefficient $C=0.65$ stabilizes the left flank even for higher average degrees $k$
	(this case is taken from Fig.\ \ref{SingleParameterScan}).
	(d) With a higher threshold $h=10$ even a lower clustering coefficient $C=0.25$ can have a stable left flank.
	The distance between the critical points shrinks for higher thresholds and both critical points are also moved
	to higher $F_+$. 
	(e) For EIN networks, a higher clustering coefficient ($C=0.5$) lowers the network's
	average sensitivity, leading this configuration to only barely pass above $\lambda=1$ between the critical points.
	From the shape of the left critical line from (b) to (d), it can also be seen that the left critical line is merely
	a continuation of the horizontal line from Fig.\ \ref{ParamScanApprox} folded upwards.}
	\label{ParameterScans}
\end{figure*}

The influence of thresholds and clustering coefficients, as well as the difference between 
EIE and EIN networks is shown in Figs.\ \ref{SecondCritPoint} (b) and \ref{ParameterScans}.
\\
So far, our networks had degree distributions centered around an average value; however,
random or Watts-Strogatz models rarely describe real-life networks.
Scale-free or similar networks with a broad degree distribution are significantly more abundant in nature.
In fact, for neuronal networks, cumulative degree distributions ranging from power laws
\cite{Varshney2011, vandenHeuvel2008, Eguiluz2005} over exponentially truncated power laws
\cite{Hayasaka2010, He2007, IturriaMedina2008, Achard2006, Gong2009} to exponential laws \cite{Modha2010,Amaral2000,Hagmann2008,SantosSierra2014} have been found, with the observation that
distributions following exponentially truncated power laws increasingly resemble true power laws
for measurements on finer scales \cite{Hayasaka2010}.
\\
In analogy to the brain, we focus on EIN networks with a broad link distribution.
For generating the topology, we 
require an algorithm that (1) can produce a scale-free graph in which low-degree nodes can exist, (2) can initialize large networks fast, (3) can produce networks with variable clustering coefficient, as we have already seen that this can have a large impact on criticality, and if possible (4) can also produce other degree distributions similar to scale-free graphs.\\
For this purpose, we adapt the algorithm described by Lo {\em et al.}\ \cite{Lo2012},
a particularly efficient implementation of preferential attachment \cite{Barabasi1999}, 
to fit our criteria.\\
In our algorithm, we start with a single node and iteratively add a connection between two nodes every two time steps, so that the sum of in and out degrees in the network increases by one per time step.
The origins and targets of these added nodes are chosen by preferential attachment, meaning that the probability of a node being chosen is proportional to the sum of its in and out degree plus an offset $\delta$, which ensures that the probability of previously unconnected nodes receiving connections is nonzero.
Further, every $m$ time steps, a new node is added to the network.
One significant difference between our algorithm and other algorithms creating scale-free graphs is that the newly added edges need not connect to the newly added node, but can instead connect any two nodes in the system, allowing low-degree nodes to exist in the final network.\\
This process is repeated multiple times and the connections of every initialization are added together into one network until the desired average degree is reached.
Finally, we add $i$ random incoming and outgoing connections to every node, where $i$ is the first integer with $i>h$, so that all nodes have the chance of being activated.
For a detailed description of this algorithm, see the Supplemental Material \cite{SuppMat}.\\
The two parameters $\delta$ and $m$ control whether the resulting degree distribution is scale free or an exponentially truncated power law and also the clustering coefficient.
In general, lower $\delta$ and higher $m$ lead to scale-free distributions with high clustering, whereas high $\delta$ and low $m$ lead to low clustering truncated power law distributions.\\

\begin{figure}
\includegraphics[width=1\linewidth]{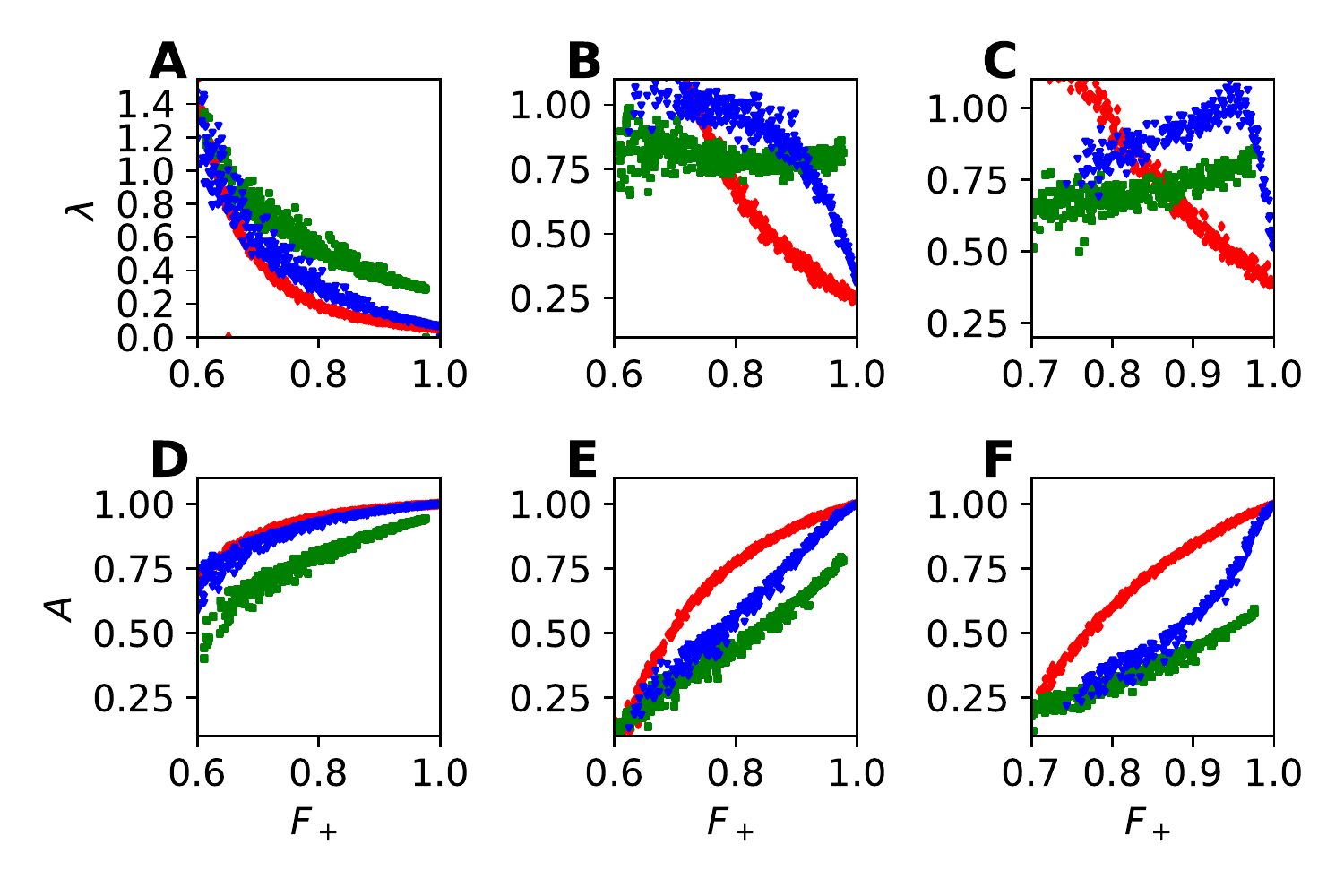}
\caption{(a)--(c): Sensitivity $\lambda$ and (d)--(f): Activity $A$ as a function of $F_+$ for an
	exponentially truncated power law network with low clustering coefficient $(\delta=40,m=2)$(red) and a
scale-free network with high clustering coefficient, where the largest node is
excitatory $(\delta=1, m=10)$(green) or inhibitory (blue) at $k=40$, $N=10^4$, and (a),(d): $h=1$, (b),(e): $h=7$, (c),(f): $h=10$.
The sensitivity and activity for the highly clustered network were measured as the average within the network's attractor.}
\label{SensitivitySplit}
\end{figure}

Studying the dynamics of EIN networks with such a topology, 
we find that for scale-free graphs the right critical point still exists (see Fig.\ \ref{SensitivitySplit}),
and that the sensitivity $\lambda$ splits into two paths on the right flank and is therefore no longer
solely dependent on $F_+$.
The two different paths are dependent on whether the network's largest node is excitatory or inhibitory
(in our algorithm, there is a clear hierarchy between nodes, dictated by when they were introduced to the network,
and therefore the first node is always clearly larger than the rest, so no multiple nodes are competing
for the spot of largest node).
Similarly to the existence of the left flank in WS networks, this split in the sensitivity is amplified
by high clustering coefficients and thresholds.
\\
Figure \ref{SensitivitySplit} also shows the existence of the left flank's second critical point for
high clustering coefficients $C$ and thresholds $h$; see Fig.\ \ref{SensitivitySplit}.
For low clustering coefficients, the left flank still dies out.
High clustering coefficients and thresholds lower the sensitivity curve's slope,
so that for certain parameters the sensitivity, and therefore criticality,
is almost constant over a wide area of $F_+$; see Fig. \ref{SensitivitySplit} (b).

To summarize, in threshold neural networks, a phase transition between
a chaotic and a quiescent regime has been found for highly clustered networks
with exclusively excitatory-inhibitory nodes.
This critical point exhibits a persisting, yet low level of average activity
(which in unclustered networks would die out).
Besides the requirement of a certain level of clustering, it is robust both
for random as well as broad (scale-free) degree distributions.
\\
This new critical point is of particular interest to neuroscience because
it is relatively independent of the degree $k$
and may, therefore, occur at the large average degree present in brains.
Furthermore, the main prerequisites for this critical point's existence are present in the brain:
a highly clustered architecture and nodes being either exclusively excitatory or inhibitory (Dale's principle).
\\
It can only be speculated what role criticality may play in nature.
It has been discussed that it could optimize a network's information processing capabilities.
Yet also, dynamical phase transitions are a simple means that physics provides, allowing
a complex system to tune to an intermediate activity regime with great ease.
\\
Last but not least, research has shown that the balance between excitation and inhibition in the brain,
which needs to be a specific value for a network to be critical in our model, is vital for a functioning brain
\cite{Haider2006, Rubenstein2003, Davenport2019, Antoine2019, Murray2018}
and that disturbing this balance can negatively impact information processing \cite{Yizhar2011}.
Interestingly, the ratio of excitatory and inhibitory neurons in brain networks
is observed to be almost constant throughout an organism's development,
and feedback algorithms that regulate this ratio are currently discussed \cite{Sahara2012}. 
This supports our hypothesis that the critical point described in this paper,
resulting from the statistical mechanics of a dynamical phase transition, 
may provide a natural target value for mechanisms that regulate the 
excitation-inhibition balance in the brain.

\bibliography{references}
\end{document}


\title{ Supplemental Material to: Critical excitation/inhibition balance in dense neural networks}

\author{Lorenz Baumgarten}
\email[]{lbaumgarten@itp.uni-bremen.de}

\author{Stefan Bornholdt}
\email[]{bornholdt@itp.uni-bremen.de}

\affiliation{Institut f\"ur Theoretische Physik, Universit\"at Bremen, 28759 Bremen, Germany}

\date{\today}
\pacs{}
\maketitle

\section{annealed approximation}
\label{AnnApp_Appendix}
We assume that all nodes are active, which is close to true at the critical point, and that the degree distribution is narrow enough that it can be approximated by a single peak at $k$.
Consider a single node with a fixed number $k$ of incoming signals. The probability of any specific signal $Z$ is
\begin{align}
p(Z) = \binom k {k_+} F_+^{k_+}F_-^{k_-},
\end{align}
where $F_-=1-F_+$ and $k_+$ and $k_-$ are the numbers of incoming excitatory and inhibitory signals, respectively.
We want to calculate the probability $p_\text s$ that the node will change its state in the next timestep if one of the incoming signals is turned off.
This can only happen if the sum of incoming signals is
\begin{align}
	k_+-k_-&=h\\
\text{or } k_+-k_-&=h+1.
\end{align}
Note that for given $k$ and $h$ only one of these two cases can occur since $k$ and $h$ need to both be odd or both be even for the first case and have different parities for the second case.\\
In the first case, an inhibitory incoming signal needs to be turned off to effect a flip.
Therefore, the fraction of connections whose disabling would produce a flip is
\begin{align}
\frac{k_-}k = \frac {k-h}{2k},
\end{align}
and for the second case, where an excitatory connection has to be turned off to effect a flip, the fraction is
\begin{align}
\frac{k_+}k = \frac {k+h+1}{2k}.
\end{align}
The respective damage spreading probabilities are
\begin{align}
p_\text s ^{(-)} 	&= \frac {\sum_{Z\in Z_{(-)}} p(Z)}{\sum_Z p(Z)} \frac {k-h}{2k}\nonumber\\
			& = \binom k {\frac{k+h}2} F_+^{\frac {k+h} 2}F_-^{\frac{k-h}2} \frac {k-h}{2k},\\
p_\text s ^{(+)} 	&= \frac {\sum_{Z\in Z_{(+)}} p(Z)}{\sum_Z p(Z)} \frac {k+h+1}{2k}\nonumber\\
			&= \binom k {\frac{k+h+1}2} F_+^{\frac {k+h+1} 2}F_-^{\frac{k-h-1}2} \frac {k+h+1}{2k}.
\label{p_s^+(k)}
\end{align}
Because we assume high activity, which requires $k_+-k_->h$, equation (\ref{p_s^+(k)}) is used in the main text.
For $k\gg h$ and $k\gg1$, the probabilities can be approximated by
\begin{align}
p_\text s^{(-)}\approx p_\text s^{(+)}\approx p_\text s\approx \frac 12 \binom k {\frac k2} \left(F_+F_-\right)^\frac k2.
\end{align}
A sensitivity of $\lambda=1$ is equivalent to a damage-spreading probability of
\begin{align}
p_\text s = \frac 1 k.
\end{align}
We use Stirling's approximation on the binomial coefficient in $p_\text s$ to arrive at
\begin{align}
\frac 1k &= \frac {2^k}k \sqrt{\frac k {2\pi}} \left(F_+F_-\right)^\frac k2\nonumber\\
\Leftrightarrow 0 &= F_+^2 - F_+ + \frac 14 \left(\frac {2\pi}k\right)^\frac 1k\nonumber\\
\Rightarrow F_+ &= \frac 12 \left[1\pm\left\{1-\left(\frac{2\pi}k\right)^\frac 1k\right\}^\frac 12\right].
\end{align}
Only the solution with $F_+>0.5$ is realistic, since the assumption of high activity does not hold for $F_+<0.5$.

\section{Methods}
\subsection{Measuring avalanche sizes and lengths}
\label{Methods:Avalanches}
We initialize random networks with random activity and let their dynamics evolve until the network runs into an attractor.
If the average sensitivity within the attractor is within a small margin of one ($1-\delta < \lambda < 1+\delta$ with $\delta=10^{-2}$), we change the state of one node and again simulate the network dynamics until the network either returns to its old attractor or reaches a new one.
During this procedure, we count the number of nodes that had a different state than they would have had in the untouched initial attractor at the corresponding time, and the number of timesteps it takes to arrive at an attractor.
This is repeated for all nodes in the network.

\subsection{Existence of the second critical point}
\label{Methods::SecondPoint}
For large clustering coefficients and high thresholds, transient lengths can become exceedingly long on the left flank, and therefore the assertion in the main text that letting networks evolve for $2\cdot10^3$ timesteps before measuring is sufficient to probe the network's dynamics may not always be correct here.
Some left flanks may seemingly exist, but will eventually die out after a long time.
Although the measurements shown in the main text were, unless otherwise stated, not explicitly done after the transient period, because waiting until the network reaches an attractor is not feasible for large sensitivities, the existence of all critical points shown was verified by letting the network reach an attractor and measuring the average sensitivity within the attractor.\\
Further,  we also tested whether this left flank was merely an aberration caused by the WS model's unrealistic ring structure by repeating our measurements for a two-dimensional network in which node position's were randomly chosen within a square area and node connections were established according to a decreasing exponential probability function of node distance (with periodic boundary conditions).
By varying this probability function, different clustering coefficients could be achieved. We verified that the left flank, and therefore also the second critical point, do exist in these more realistic networks.

\subsection{Scale-free and exponentially truncated scale-free graphs}
\label{Methods::ScaleFree}
\begin{figure}
	\includegraphics[width=1\linewidth]{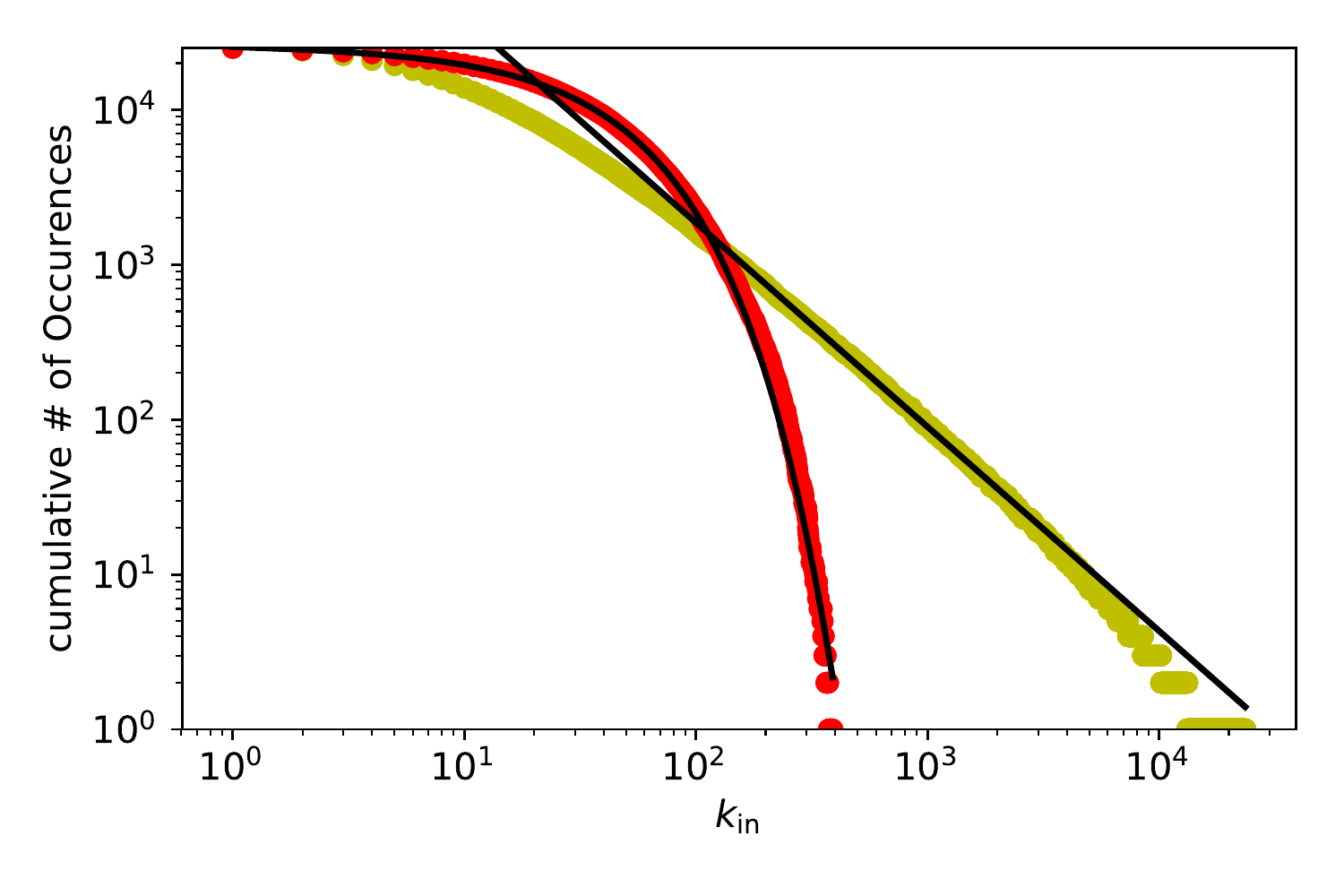}
\caption{Cumulative in-degree distributions for a scale-free graph with $\delta=1$, $m=10$, $k=40$, $N=2.5\cdot10^4$, $h=0$, and $C\approx0.52$ (yellow) and  an exponentially truncated scale-free graph with $\delta=40$, $m=2$, $k=40$, $N=2.5\cdot10^4$, $h=0$, and $C\approx0.005$ (red). Solid black lines show power law and exponentially truncated power law fits, respectively. The power law's slope is -1.3.}
\label{DegDis}
\end{figure}

For our preferential attachment, we do not differentiate between a node's in- and outdegree, i.e. the probability of a node being chosen via preferential attachment is proportional to the sum of a node's in- and outdegree, and the probability of a node being chosen as the origin or the target of a node are also equal.
Therefore, for simplicity, we refer to the sum of in- and outdegree simply as the degree in this section, as if the network were undirected.\\
We start with a single node  and add one edge between two nodes, both the edge's origin and target chosen by preferential attachment, in every two timesteps, so that the network's total degree increases by one per timestep.
After every $2m$ timesteps, we also add one node to the system.
Ignore for now that it is not always possible to add edges to the system for small $t$.
We do not, however, use the nodes' actual degrees to choose which nodes will gain an edge, but instead calculate expected degrees for the nodes.
The expected degree at timestep $t_j$ of a node $v_k$ that was introduced to the network at time $t_k$ is
\begin{align}
Exp&Deg(t_j, v_k) \nonumber\\
	&= ExpDeg(t_{j-1},v_k) + \frac{ ExpDeg(t_{j-1},v_k) } { total(t_{j-1}) } \nonumber \\
	&= ExpDeg(t_{j-1},v_k)(1+1/total(t_{j-1})) \nonumber \\
	&= ExpDeg(t_k,v_k)\prod_{i=k}^{j-1}(1+1/total(t_i)),
\end{align}
where $total(t_j)$ is what we call the system's total adjusted degree at time $t_j$.
In our definition of a node's adjusted degree (and the expected degree), we include a bias $\delta$ that is added to a node's number of edges.
The total adjusted degree is therefore
\begin{align}
total(t_j) = t_j + \delta n_j,
\end{align}
where $n_j$ is the number of nodes in the system at time $t_j$.
This bias is necessary because all nodes start with zero edges and would therefore never receive any new edges for $\delta=0$.
Using the nodes' expected instead of their actual degrees enables us to parallelize our algorithm, significantly speeding up a network's initialization.
Note that for the expected degree the issue of being unable to add any edges for small network sizes is irrelevant.
We simply pretend that nodes had the degree they could have, had we actually added an edge in every two timesteps, which still simulates preferential attachment. This simplistic approach enables us to simplify products in the following calculations which would otherwise be computationally expensive.\\
More useful than  every single node's degree, however, is the cumulative degree
\begin{align}
ECu&mDeg(t_j,v_k) \nonumber \\
	&=\sum_{i=1}^k ExpDeg(t_j,v_i) \nonumber \\
	&=\sum_{i=1}^k ExpDeg(t_k,v_i) \prod_{i=k}^{j-1}(1+1/total(t_i)) \nonumber \\
	&=total(t_k)\prod_{i=k}^{j-1}(1+1/total(t_i)) \nonumber \\
	&=total(t_k) \cdot \xi(t_k, t_j),
\end{align}
as it allows us to do a fast binary search to find the node an edge needs to be connected to, without having to calculate and sum up expected degrees for a large number of nodes.
The factor $\xi(t_k, t_j)$ can easily be calculated when the number of nodes currently in the network $n$ is constant
\begin{align}
\xi(t_k,t_j) 	&= \prod_{i=k}^{j-1} \left( 1+ \frac 1 {i + \delta n} \right) \nonumber \\
		&= \left( \frac{k+\delta n + 1}{k+\delta n}\right) \left( \frac{k+\delta n + 2}{k+\delta n+1}\right) \dotsb \left( \frac{j+\delta n}{j+\delta n-1}\right) \nonumber \\
		&= \frac {j + \delta n} {k + \delta n}.
\end{align}
To calculate $\xi(t_k,t_j)$ even if nodes are added to the system between $t_k$ and $t_j$, we split the product into parts with constant $n$
\begin{align}
\xi(t_k, t_j) 	= &\frac {k + 2m + \delta n_k}{k + \delta n_k} \times \frac{ k + 4m + \delta (n_k+1)}{ k + 2m + \delta n_k} \nonumber \\
		&\times \dotsb \times \frac{ k + 2m(n_j-n_k) + \delta n_j}{ k + 2m(n_j-n_k-1) + \delta (n_j-1)} \nonumber \\
		&\times \frac{ j +\delta n_j }{ k+2m(n_j-n_k) + \delta n_j },
\end{align}
The product in the first two lines can be written as
\begin{align}
&\prod_{i=0}^{n_j-n_k-1} \frac{ k + 2mi + 2m + \delta(n_k+i)} { k+2mi + \delta(n_k+i)} \nonumber \\
&=\prod_{i=0}^{n_j-n_k-1} 1 + \frac 1 {c + i(1+\delta/2m)} \nonumber \\
&=\frac { \Gamma(\frac c d) \Gamma( b + \frac c d + \frac 1 d) } { \Gamma(\frac c d + \frac 1 d) \Gamma(b + \frac c d)}, \\
&\text{with } b=n_j-n_k,~c = (k+\delta n_k)/2m,~d=1+\delta/2m. \nonumber
\end{align}
The cumulative degree is then
\begin{align}
E&CumDeg(t_j,v_k) = total(t_k) \nonumber \\
&\times
	\frac { \Gamma(\frac c d) \Gamma( b + \frac c d + \frac 1 d) } { \Gamma(\frac c d + \frac 1 d) \Gamma(b + \frac c d)} \times \frac{ j +\delta n_j }{ k+2m(n_j-n_k) + \delta n_j }.
\end{align}
At time $t_j$, a node can now be chosen via preferential attachment by choosing a random number $\eta$ between zero and one, and finding the first $v_k$ for which $ECumDeg(t_j,v_k)/total(t_j) \ge \eta$.\\
It is likely that edges between the largest nodes will be added multiple times during a network's initialization.
As it would be computationally expensive to check whether an edge already exists in the system, we ignore additional edges and remove them after initialization so weights remain as $\pm 1$.
This, and also the inability to add an edge every two timesteps for low $t$, leads to the network's eventual average degree being unpredictable.
Therefore, we repeat the initialization process, adding up all of the single initializations' edges, until $k\approx k^*$, where $k$ is the network's average degree and $k^*$ is the desired average degree.
Finally, we add edges from $i$ permutations of the network's nodes to the unpermutated nodes, with $i$ being the smallest integer with $i>h$, to ensure that every node has a chance of being activated and participating in the network's dynamics.\\
Our algorithm has two parameters, $\delta$ and $m$, that enable us to tune the degree distribution as well as the clustering coefficient.
Generally, lower $\delta$ and higher $m$ lead to higher clustering coefficients and scale-free degree distributions, whereas higher $\delta$ and lower $m$ lead to low clustering coefficients and exponentially truncated power law degree distributions, see FIG. \ref{DegDis}.\\
